\documentstyle[%
,multicol,epsfig,aps,prb]{revtex}

\begin{document}
 \ifpreprintsty
 \def\multb{ }
 \def\multe{ }
 \else
 \def\multb{ \begin{multicols}{2}}
 \def\multe{ \end{multicols}}
 \fi
\title{Theoretical search for Chevrel phase based thermoelectric materials}
\author{R.~W.~Nunes$^{1,2}$\cite{rn}%
, I.~I.~Mazin$^{1,2}$, and D.~J.~Singh$^1$}
\address{$^1$Complex System Theory Branch, Naval Research Laboratory,
Washington DC, 20375-53459\\
$^2$Computational Sciences and Informatics, George Mason University,
Fairfax, Virginia}
\date{\today}
\maketitle

\begin{abstract}
We investigate the thermoelectric properties of  some
semiconducting Chevrel phases.
Band structure calculations are used to compute
thermopowers and to 
estimate of the effects
of alloying and disorder on carrier mobility.
Alloying on the Mo site with transition metals like Re, Ru or 
Tc to reach a semiconducting composition
 causes large changes in the electronic structure at the Fermi 
level. Such alloys are expected to have low carrier mobilities.
 Filling with transition
metals was  also found to be incompatible with high thermoelectric
performance based on the calculated electronic structures.
 Filling with Zn, Cu, and especially with Li was found
to be favorable. The calculated electronic structures of these
filled Chevrel phases are consistent with low scattering of carriers
by defects associated with the filling.
We expect 
good mobility and high thermopower in materials with the composition
close to (Li,Cu)$_4$Mo$_6$Se$_8$, particularly when Li-rich,
and recommend this system for experimental
investigation.

\end{abstract}

 \multb
Widespread use of thermoelectric materials, as an alternative technology for
power generation and refrigeration, remains a desirable but elusive goal.
The efficiency of devices based on current state-of-the-art materials (such
as PbTe, SiGe, and Bi$_{2}$Te$_{3}$/Sb$_{2}$Te$_{3}$/Bi$_{2}$Se%
$_{3}$ alloys) is low.
 This largely restricts them to the applications where
reliability outweighs efficiency or small device sizes are needed. In
recent years, the center of gravity of the search for new thermoelectric
materials has moved towards crystallographically more complicated compounds.
This is motivated in part by the fact that low values of thermal
conductivity, needed for thermoelectric performance, are more likely in such
materials, and partly because such structures provide more avenues for chemical optimization. Good examples are 
high figure of merit ($ZT$) skutterudite
 and Zn-Sb thermoelectrics. First 
principles calculations are proving to be a useful tool in sorting out the mechanisms by which high $ZT$ values can arise in these complex materials and
in exploring the effects of various chemical modifications on the 
thermoelectric properties.
~\cite{mahan,singh-p,singh-m,singh-k}.

A good thermoelectric material has high thermopower, a low
lattice contribution to the thermal conductivity $\kappa$
and a sizeable electric conductivity\cite{mahan,slack}.
The thermoelectric figure of merit is $%
ZT = \sigma S^2 T/\kappa$, where $\sigma$ is the electrical conductivity, $S$
is the Seebeck coefficient (thermopower), and $\kappa = \kappa_{el} +
\kappa_{latt}$ is the thermal conductivity ($\kappa_{el}$
and $\kappa_{latt}$ are 
 the electronic and lattice contributions, respectively).
Current thermoelectric materials have $ZT \approx 1$.
 Large
values of $S$ are typical of doped semiconductors, while $\sigma$ is large
in metals. Starting from a material with low $\kappa$, the above expression
suggests that enhanced values of $ZT$ can be obtained by searching for
compositions which maximize the power factor, $\sigma S^2$, with respect to
the carrier concentration $n$.
An additional requirement arises from
 the Wiedemann-Franz law, which sets a
rough lower bound on the Lorentz number $L=\kappa/\sigma T$ [$L_{min}=(\pi^2
k_B/3e)$]. A value $S\geq$160 $\mu$V/K is thus needed for $ZT\geq$1,
even if $\kappa_{\rm latt}$ is negligibly small.
Empirical evidence and minimum thermal conductivity theories imply that
materials having large numbers of atoms in the unit cell
and large atomic masses (soft phonons),
are most likely to have the requisite low
thermal conductivities.~\cite{mahan}
Here we investigate the
materials in the family of molybdenum (Mo)
cluster compounds known as ``Chevrel phases''. These are based on the
binaries Mo$_{6}$X$_{8}$, where X is a chalcogen
(S, Se, Te). The crystal structure contains large voids,
which may be filled to yield a large variety of ternary
compounds with the general formula M$_{x}$Mo$_{6}$X$_{8}$, where M can be a
simple or transition metal atom, or a rare-earth element. This
provides opportunities for obtaining low thermal conductivities, analogous
to skutterudites, as well as a chemical knob for modifying the electronic
properties. These
materials fulfill the requirements for low $\kappa $ outlined above, and 
experimental evidence indicates that their thermal conductivity is indeed
low.~\cite{jpl}

The band structure of  the unfilled Mo$_{6}$X$_{8}$ is
 relatively simple,
 despite the crystallographic complexity.
 Following Ref.\onlinecite
{nohl}, the band structure
can be described as deriving from a collection of pseudocubic
Mo$_{6}$X$_{8}$ clusters. The chalcogens form a distorted cube, X$_{8}$, and
the Mo's occupy the face centers of the cube.  The on-site energies of the
X-$p$ and Mo-$d$ orbitals are close; for instance, in Mo$_{6}$Se$_{8}$
$E_d-E_p\approx 0.2-0.4$ eV (in Mo$_{6}$S$_{8}$ it is larger, about
0.9 eV). These orbitals are strongly hybridized:
the hopping amplitude $t_{pd\sigma}$ is of the order of 2 eV. Two out of
five $d$-orbitals for each Mo atom are affected by the ${pd\sigma}$
bonding (for the $xy$ face they are $xy$ and $3z^2-r^2$). The total 
number of affected states is $8\times 3+6\times 2=36$, of which there are
12 non bonding $p$ states, 12 bonding, and 12 antibonding ${pd\sigma}$
states. In unfilled Mo$_{6}$X$_{8}$, 24 bonding and nonbonding states
(48 in both spin channels)
are all occupied, and the antibonding states are much higher and
are never populated in the real compounds. The remaining Mo $d$
orbitals form 18 bands sitting inside the gap between the non-bonding
and anti-bonding ${pd\sigma}$ bands. The $dd\sigma$ hopping amplitude
is also very large, $\approx 1.5$ eV, and as a result these 18 bands
also  form bonding and antibonding combinations (there are no 
nonbonding states for $d$-orbitals on an octahedron). It can be shown that
the number of bonding states is 12, and of antibonding states 6, so there
is a gap between 36th and 37th state in the Mo$_{6}$X$_{8}$ cluster.

The total number of electrons 
in the $pd$ states in Mo$_{6}$X$_{8}$ is $8\times 4+6\times 6=68$,
4 electrons short of reaching the above mentioned gap. Thus, one conjectures
that doping 4 electrons in the system will make it semiconducting,
and that the states of the bottom of the gap would be 
predominantly Mo $d$. In particular, one expects the bands  below the gap
to be sensitive to substitution on the metal site, but not to 
filling the voids. However, the desired doping requires substantial
filling, {\it e.g.} with
 4 monovalent atoms (Li, Cu), or 2 divalent ones (Zn) per formula unit.

In our calculations, the lithium (Li) filled selenide, Li$_{4-\delta
}$Mo$_{6}$%
Se$_{8}$ is found to be particularly favorable for thermoelectric
applications from an electronic point of view. Other favorable choices 
include Cu$_{4-\delta}$Mo$_{6}$Se$_{8}$ and Zn$_{2-\delta}$Mo$_{6}$Se$_{8}$,
while transition metal fillers or alloying on the Mo site
are found to have too low carrier mobilities to be candidates for
thermoelectric applications.\cite{note} We concentrate on the Se based
compounds, as opposed to more common sulfides.
Se has higher mass, favorable to low lattice heat
conductance. Furthermore, selenides tend to be more
covalent,\cite{yvon} which is better from the point of view of carrier
mobility.
Experimentally there is a large homogeneity region in the
Li$_x$Mo$_{6}$Se$_{8}$ pseudobinary  phase diagram near this
composition.~\cite{selwyn}

We use a density-functional approach within the
local-density approximation (LDA) for the exchange and correlation
to calculate the electronic structures.
In accord with considerations presented above, Li$_{4}$Mo$_{6}$Se$_{8}$ 
was indeed found to be
a semiconductor with an LDA gap of about 1.4 eV.
For most calculations we used
the general potential linearized augmented plane wave
(LAPW) method~\cite{oka,singh1} with local-orbital
extensions\cite{singh2} to relax  linearization errors
and treat  the $4p$ semi-core states of Mo on an equal footing
with the valence states. We employ a well-converged basis set containing $%
\sim $2000 basis functions, with muffin-tin sphere radii
 of 2.25 a.u. for Mo
and Se, and of 2.0 a.u. for Li. The calculations were based
on the reported crystal structure of
Li$_{3.2}$Mo$_{6}$Se$_{8}$.~\cite{cava} In the present work, we do not
take into account the trigonal distortion that was reported for
$\delta<0.4$ in Ref.\onlinecite{dahn}. This is probably associated with
partial ordering of the Li ions.

We also considered several other dopants to the Mo$_{6}$Se$_{8}$ phase, to
add four electrons per formula unit and make it a semiconductor.
These include doping with zinc, copper, and some transition metals at the
interstitial sites, as well as substitution of ruthenium, technetium, and
rhenium on the metal (Mo) site. Zn and Cu fillings were found to be
favorable, but the Chevrel phases with the needed concentration have not been
synthesized experimentally. Transition metal substitutions were found to be
unfavorable due to strong carrier scattering (as discussed above, the states
below the gap are predominantly of  Mo-Mo bonding character, thus
substitution at the Mo site has a drastic effect on these states).
A detailed account of these
studies will be published elsewhere.

\begin{figure}[tbp]
\centerline{\epsfig{file=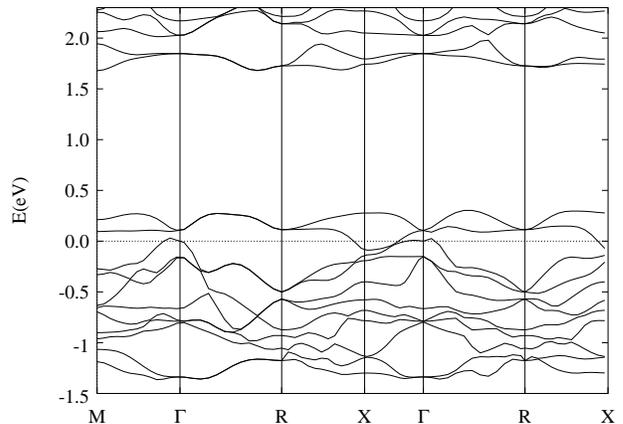,width=0.75\linewidth,angle=-90}}
\vspace{0.1in} \setlength{\columnwidth}{3.5in}\vspace{0.1in}    
\caption{Band structure of the Mo$_6$Se$_8$}
\label{fig1}      \end{figure}
Transport properties were obtained from the calculated LAPW electronic
 structures using
kinetic  (Bloch-Boltzmann) transport theory.~\cite{ziman}
The only further approximation was that of an isotropic and energy
independent relaxation time for electrons. Unless there is particularly 
sharp structure of the density of states near the Fermi level ({\it
e.g.} Pd or heavy fermion materials), this approximation
is quite good.

The structure of Chevrel compounds  can be described
as a collection of the above mentioned pseudocubic Mo$_6$X$_8$ clusters
with a small symmetry lowering rhombohedral distortion%
~\cite{yvon,nohl}. Several interconnecting interstices, which are empty
sites in the binary phases, form channels along the three rhombohedral
directions.
A large variety of ternary compounds may be
synthesized by filling these empty sites, {\it e.g.} Li$_{x}$Mo$%
_{6}$Se$_{8}$, and also the
prototype large-cation compound
PbMo$_{6}$S$_{8}$, which is obtained by placing Pb into the
largest interstitial site at the origin of the rhombohedral unit cell.
PbMo$_{6}$S$_{8}$ and similar
 materials have been extensively studied due to their interesting
superconducting properties,\cite{yvon,nohl,chevrel} but semiconducting
compositions have  attracted less interest. Unlike large cations, like Pb,
small cations such as Li are distributed over 12 sites arranged
as two sets of concentric sixfold rings surrounding the large interstice at
the origin of the rhombohedral cell.\cite{cava,ritter} For intermediate
concentrations ($1<x<4$), the relative occupancy of these two
rings has been a subject of debate, at least in the case of the sulfide.\cite
{cava,ritter} However, for $x=4$ it is commonly accepted that three Li atoms
occupy the outer ring, with the fourth Li atom occupying the inner ring.
There is a tendency for Li atoms to order partially, however, this ordering
is very little studied.

Bonding in these compounds is understood to derive from covalent bonds
between the Mo atoms in the octahedral cluster, and between Mo and X
(Ref. \onlinecite{nohl}). The latter bond has mixed
ionic/covalent character. As mentioned, the Mo-Se bonds are
 less ionic than Mo-S
bonds.~\cite{yvon} The valence bands of Mo$_{6}$Se$_{8}$ shown in Fig~\ref
{fig1} are thus composed of covalent Mo-Mo and Mo-Se bonds.
The
conduction band is partially occupied (four holes per unit cell) at
the top of the valence band manifold. As mentioned, the top states
in this band consist primarily of bonding Mo-Mo
states along the edges of the Mo octahedra.

\begin{figure}[tbp]
\centerline{\epsfig{file=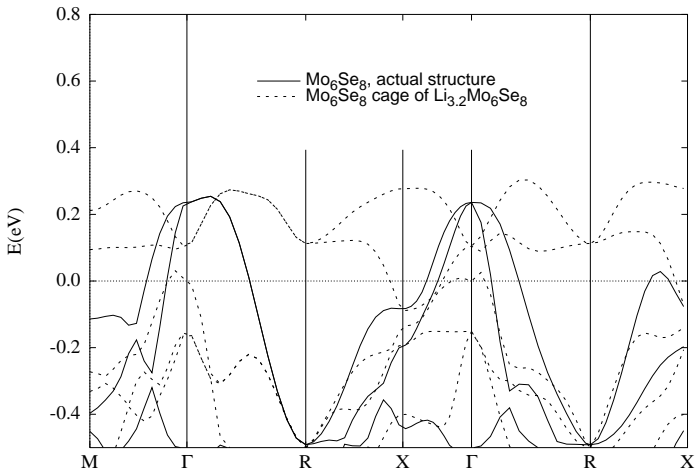,width=0.75\linewidth,angle=-90}}
\vspace{0.1in} \setlength{\columnwidth}{3.5in}\vspace{0.1in}    
\caption{Band structure of the Mo$_6$Se$_8$ and of the same compound 
with the Mo and Se atoms in their position in
Li$_{3.2}$Mo$_6$Se$_8$ (but without
Li). Note the large changes due to relatively small changes in the Mo-Se
and Mo-Mo bond lengths, and overall cell geometry.}
\label{fig2} \end{figure}
In addition to providing the four extra electrons needed to dope the
compound to the semiconducting regime, the insertion of Li in the
interstitial sites leads to changes in the structure of the Mo-Se frame.{\it 
e.g.}, upon saturation of the Mo-Mo bonds, the octahedron contracts and
becomes more regular, and the Mo-Se distances increase. The 
corresponding change in electronic structure is shown
 in Fig~\ref{fig2},
where the bands of Mo$_{6}$Se$_{8}$ near the valence band edge
are compared for two crystal structures: the experimental structure of
the binary system and
the structure of the Mo-Se frame in Li$_{3.2}$Mo$_{6}$Se$_{8}$ (but
without Li). Fig.~\ref{fig2} indicates that the changes in the Mo-Se frame
lead to an enhancement of the band masses. Moreover, the valence band maxima
are shifted to general k-points, which implies that a larger number of
valleys contribute to the transport integrals, enhancing $\sigma$
without degrading $S$.

\begin{figure}[tbp]
\centerline{\epsfig{file=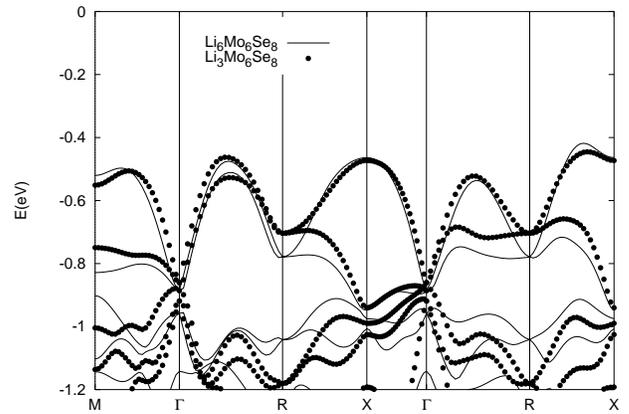,width=0.95\linewidth}}
\vspace{0.1in} \setlength{\columnwidth}{3.5in}\vspace{0.1in}    
\caption{Band structure of the ordered rhombohedral Li$_3$Mo$_6$Se$_8$
compared with that of a hypothetical rhombohedral Li$_6$Mo$_6$Se$_8$.}
\label{fig3} \end{figure}
As mentioned, we use the experimental
structure obtained for Li$_{3.2}$Mo$_{6}$Se$_{8}$ by Cava {\it et al.}
 in our band-structure calculations of the Li$_{3}$Mo$_{6}$Se%
$_{8}$ compound, with the three Li atoms  placed in the outer
interstitial ring, preserving the threefold rotation
symmetry. This band-structure is then used to compute the
transport integrals for the semiconducting composition, under the assumption
that the system follows a rigid band behavior as we fill the valence bands.
In order to test this assumption, we also computed the band structure of Li$%
_{6}$Mo$_{6}$Se$_{8}$. In this case, we placed the three additional Li atoms
in the inner rings, once again keeping the threefold rotation about the
trigonal axis. The results are shown in Fig.~\ref{fig3}. The primary
contribution to the transport properties is derived from the top of the
valence band, which occurs along the R-X line in Fig.~\ref{fig3}. The two
sets of bands are very similar near the top of the bands, despite the very
strong perturbation of the self-consistent potential, upon the introduction
of three additional Li ions and three additional electrons. 
This is key because it indicates weak scattering due to Li disorder and
vacancies, which is a prerequisite for obtaining reasonable mobilities
and thermoelectric performance. The most
pronounced difference between these two sets of bands near the top of the
bands is found along the $\Gamma $-R directions, where the eigenvalues are
shifted downwards in the Li$_{6}$Mo$_{6}$Se$_{8}$, with respect to the
band maxima. Nevertheless, given that the perturbation in this case is
substantially stronger than in the composition of interest ($x=4$), we see
Fig.~\ref{fig3} as an indication that rigid band behavior between the $x=3$
and $x=4$ compositions holds to a good approximation. The observed
rigid-band behavior also shows that the crystal potential felt by 
carriers at and near the Fermi level is only moderately affected by the
exact position of the Li ions. The carrier mobility is determined by the
electron scattering time, which in turns is defined by the change of the
crystal potential upon moving Li atoms.
Correspondingly,
when the change is small, the carrier mobility is expected to be
only weakly reduced by Coulomb scattering due to the disorder on the Li
site. The negative effect on the mobility will be even further reduced by
partial ordering of
Li atoms, which is known to occur in Li$_{4-\delta}$Mo$_{6}$Se%
$_{8}$. (A phase transition to a lower symmetry trigonal phase has been
reported for $\delta < 0.4$ in this compound.~\cite{dahn}) 
Furthermore, although the ordering processes in the mixed-filler system 
(Cu,Li)$_{4-\delta}$Mo$_{6}$Se$_{8}$ have hardly been studied at all, one
might anticipate
 that, say, CuLi$_3$Mo$_{6}$Se$_{8}$ (this compound forms), would
be better ordered than Li$_{4}$Mo$_{6}$Se$_{8}$.

\begin{figure}[tbp]
\centerline{\epsfig{file=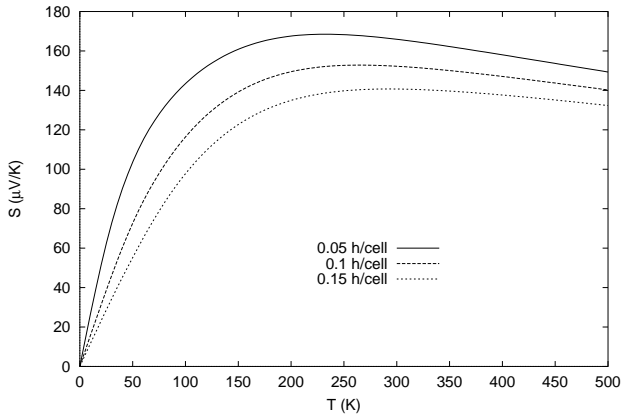,width=0.95\linewidth}}
\vspace{0.1in} \setlength{\columnwidth}{3.5in}\vspace{0.1in}    
\caption{Calculated thermopower of Li$_{4-\delta}$Se$_8$ for 
$\delta=$ 0.05, 0.1, and 0.15.}
\label{fig4}
\end{figure}
The calculated thermopower as a function of temperature is shown in Fig.~\ref
{fig4} for three different doping levels. All three curves reach a
maximum at around 200 K.
At a doping of about 0.1 hole/cell
hole/cell (corresponding to a concentration of 3$\times 10^{20}$
holes/cm$^{3}$%
) we obtain a maximum Seebeck coefficient of $\sim
150~\mu $V/K  at $T \sim 250$ K. Without 
quantitative data on the lattice thermal conductivity and carrier 
mobility it is not possible to determine $ZT$, but it is worth mentioning
that these high values at high carrier concentration, if combined with
reasonable $\kappa_{\rm latt}$, are compatible with  
 $ZT\geq 1$, when the material is optimized.

In summary, we present first principles calculations of filled Chevrel
phases based on Mo$_{6}$Se$_{8}.$ The most promising from the point of
view of potential thermoelectric applications is Li$_{4-\delta }$Mo$_{6}$Se$%
_{8}$, with $\delta \sim 0.1.$ This material can possibly
have values of $ZT$ of the order of 1 or larger.
Among the most important questions
 about this and other filled Chevrel compounds,
from the point of view of thermoelectric applications are: (1) whether the
effects neglected in the current study, namely the local distortion
of the Mo$_6$Se$_8$ cage near Li and partial ordering of Li atoms
will not adversely affect the carrier mobility (the former effect lowers
the mobility, and the latter increases it), and (2) whether the alloying
with Li or Li/Cu  would  be sufficiently effective
in scattering  heat conducting phonons.
In this regard Cu and Li filled Mo$_6$Se$_8$ are similar electronically 
but presumably very different vibrationally, so the Li/Cu ratio
may be a useful ``knob'' for controlling the thermal conductivity.
We hope that our calculations will encourage experimentalists to look
closer at this promising system, (Li,Cu)$_{4-\delta }$Mo$_{6}$Se$%
_{8}$.

The authors are thankful to Terry Caillat and other members of the JPL
thermoelectric group for many discussions, and for calling our attention to
the Chevrel phases, and to Mark Rikel for calling our attention to
 Cu and Li as potential
fillers. This work was supported by DARPA and ONR.
Computations were performed using the DoD HPCMO facilities at ASC.






\multe
\end{document}